\def\year{2022}\relax
\documentclass[letterpaper]{article} % DO NOT CHANGE THIS
\usepackage{aaai22}  % DO NOT CHANGE THIS
\usepackage{times}  % DO NOT CHANGE THIS
\usepackage{helvet}  % DO NOT CHANGE THIS
\usepackage{courier}  % DO NOT CHANGE THIS
\usepackage[hyphens]{url}  % DO NOT CHANGE THIS
\usepackage{graphicx} % DO NOT CHANGE THIS
\usepackage{multirow}
\usepackage{booktabs}
\usepackage{breqn}
\usepackage{enumitem}
\usepackage{subfig}
\usepackage{amsmath, amsthm, amssymb}
\newtheorem{prop}{Proposition}
\DeclareMathOperator*{\argmax}{arg\,max}

\usepackage{natbib}  % DO NOT CHANGE THIS AND DO NOT ADD ANY OPTIONS TO IT
\usepackage{caption} % DO NOT CHANGE THIS AND DO NOT ADD ANY OPTIONS TO IT
\DeclareCaptionStyle{ruled}{labelfont=normalfont,labelsep=colon,strut=off} % DO NOT CHANGE THIS
\usepackage{algorithm}
\usepackage{algorithmic}

\usepackage{newfloat}
\usepackage{listings}
\floatstyle{ruled}
\newfloat{listing}{tb}{lst}{}
\floatname{listing}{Listing}
\usepackage{color}
\title{DROP: Deep relocating option policy for optimal ride-hailing vehicle repositioning}

\author {
    Xinwu Qian,\textsuperscript{\rm 1}
    Shuocheng Guo, \textsuperscript{\rm 1}
    Vaneet Aggarwal \textsuperscript{\rm 2}
}
\affiliations {
    \textsuperscript{\rm 1} Department of Civil, Construction and Environmental Engineering, The University of Alabama, USA\\
    \textsuperscript{\rm 2} School of Industrial Engineering, Purdue University, USA\\
    xinwu.qian@ua.edu, sguo18@crimson.ua.edu, vaneet@purdue.edu
}

\begin{document}

\maketitle

\begin{abstract}
In a ride-hailing system, an optimal relocation of vacant vehicles can significantly reduce fleet idling time and balance the supply-demand distribution, enhancing system efficiency and promoting driver satisfaction and retention. Model-free deep reinforcement learning (DRL) has been shown to dynamically learn the relocating policy by actively interacting with the intrinsic dynamics in large-scale ride-hailing systems. However, the issues of sparse reward signals and unbalanced demand and supply distribution place critical barriers in developing effective DRL models. Conventional exploration strategy (e.g., the $\epsilon$-greedy) may barely work under such an environment because of dithering in low-demand regions distant from high-revenue regions. This study proposes the deep relocating option policy (DROP) that supervises vehicle agents to escape from oversupply areas and effectively relocate to potentially underserved areas. We propose to learn the Laplacian embedding of a time-expanded relocation graph, as an approximation representation of the system relocation policy. The embedding generates task-agnostic signals, which in combination with task-dependent signals, constitute the pseudo-reward function for generating DROPs. We present a hierarchical learning framework that trains a high-level relocation policy and a set of low-level DROPs. The effectiveness of our approach is demonstrated using a custom-built high-fidelity simulator with real-world trip record data. We report that DROP significantly improves baseline models with 15.7\% more hourly revenue and can effectively resolve the dithering issue in low-demand areas.
\end{abstract}

\section{Introduction}

The ride-hailing platform such as Uber, Lyft, and Didi is now an indispensable component of urban transportation systems, hitting 50 million daily orders by Didi~\cite{didi_50million_orders} and 5 billion orders by Uber in 2020~\cite{uber_5billion}. For the complex ride-hailing system, the optimal management of large-scale vehicle fleets relies heavily on effective order dispatching and vehicle repositioning algorithms. Specifically, order dispatching~\cite{tang2019deep,qin2020ride} is to assign open orders to available drivers aiming at minimizing customer waiting time or maximizing driver income. The vehicle repositioning~\cite{lin2018efficient,jiao2021realworld} is to proactively dispatch the idle vehicles to potential high-demand locations, which greatly saves driver's idling time so as to serve more requests in a long-term horizon. Also, a practical vehicle repositioning strategy can balance the supply-demand distribution and in turn improve the operational efficiency for the ride-hailing platform, which further depicts positive impacts on driver satisfaction and retention~\cite{jiao2021realworld}. In this study, we focus on the vehicle repositioning problem aiming to improve the efficiency of the ride-hailing system.

% which depicts great needs on an efficient vehicle relocating policy to  and balance the supply-demand gaps on a real-time scale. 
% we note that e-greedy: sample complexity is exponential, can be understood as 2d RW under hexagonal setting. vehicle dithering in low demand area.

% [dicisive supervised indepedent reward signal] primitive actions that allows the vehicles to escape the low demand areas. 
% manually design high-level relocation policy based on domain knowledge. ~\cite{xu2018large}
% cite the pre-designed actions, which is hard to acheive in a highly dynamically system.
% automatically generated options that based on state transitions produced by large-scale agents,
% we seek to learn the underlying graph laplacian, to identify the set of state with potentially high rewards. generate supervised the vehicles to escape the low demand area.

\begin{figure}[!htbp]
    \centering
    \includegraphics[width = 0.85\linewidth]{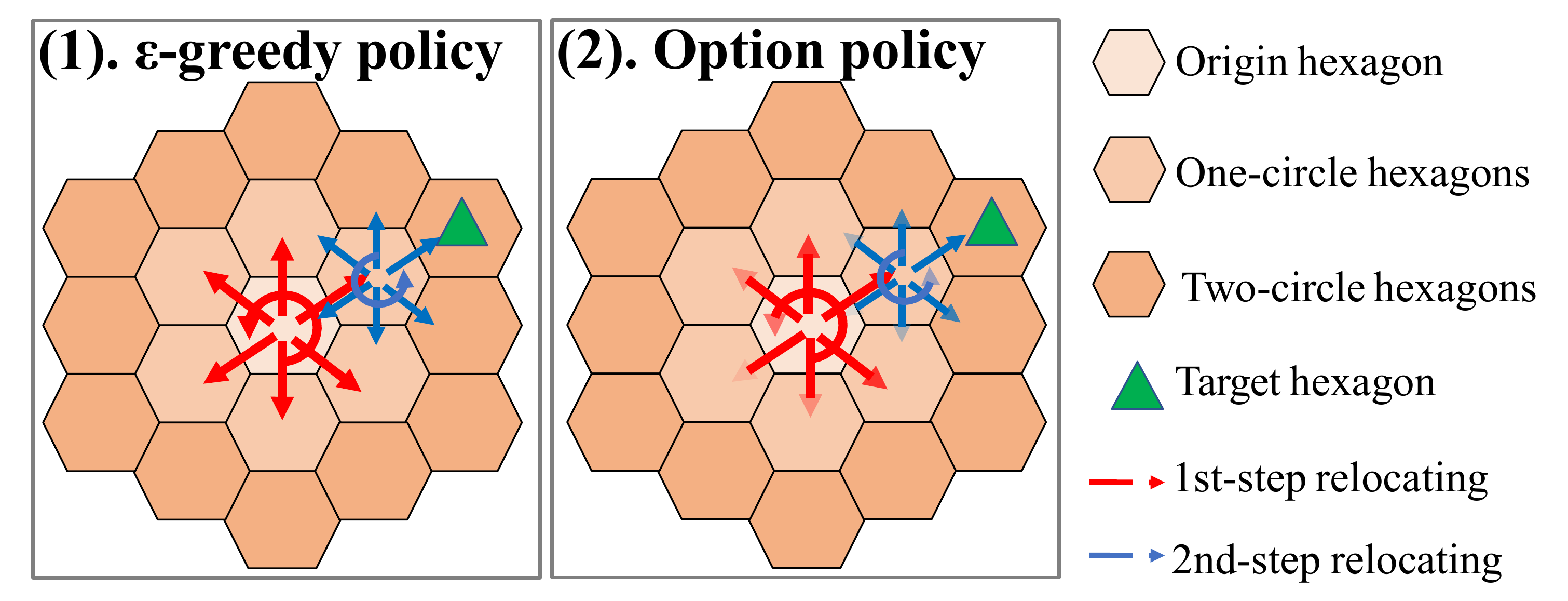}
    \caption{Illustration for relocating: (1) under $\epsilon$-greedy exploration and (2) under option policy.}
    \label{fig:reposition_prob}
\end{figure}

There were several successful applications of reinforcement learning (RL) in the vehicle repositioning problem~\cite{holler2019deep,al2019deeppool,shou2020reward, jiao2021realworld}. However, most studies perform the exploration under $\epsilon$-greedy policy on the square or hexagonal grid system, which will greatly enlarge the \textit{sample complexity (of exploration)}~\cite{kakade2003sample}. Two particular challenges  emerge under the $\epsilon$-greedy exploration for the vehicle repositioning problem. First, the $\epsilon$-greedy relocating policy  results in unbalanced exploration between low- and high-demand regions. Specifically, the low-demand regions are more occupied with repetitive relocation by the unmatched vehicles, while the active vehicles in high-demand areas are assigned to distant orders without further exploring the nearby locations. The relocation can be understood as a 2-dimension random walk process~\cite{machado2017laplacian}, where the vehicle agents are likely to get trapped without decisive guidance, thus locally dithering around their origins. As illustrated in Fig.~\ref{fig:reposition_prob}(1), the probability of touching the target depicts an exponential decay under the $\epsilon$-greedy exploration; whereas in Figure~\ref{fig:reposition_prob}(2), the agent can reach the target within fewer steps, guided by a sequence of decisive directions. 
% Besides, we note that the skewed demand distribution may also gives rise to the sample complexity issues. For example, a vehicle may serve the request targeting at the low-demand region, where the vehicle can hardly escape under the $\epsilon$-greedy policy, thus aggravating the sample complexity issue. We illustrate the `locally dithering' by considering the probability of reaching a specific range within few steps as shown in Figure~\ref{fig:reposition_prob}(the probabilities of each direction are shown as the transparency). In this regard,

In this study, we propose a deep relocating option policy (DROP) for model-free reinforcement learning models under a semi-Markov Decision Process (SMDP) setting. The SMDP can capture the long-term impacts of relocating decisions (e.g., one successful matching relies on multiple preceding relocation steps) and also potentially improve the sample complexity~\cite{brunskill2014pac}. In this regard, we design DROPs to supervise the vehicle agents relocating from oversupply to underserved regions decisively. Acknowledging that demand-supply relationship is a function of the relocation policy, we leverage the idea of graph Laplacian to approximate an embedding of the relocation policy through a constructed relocation network. The developed DROPs and the high-level relocation policy are then trained under a hierarchical framework using deep Q-networks (DQN). We summarize the major contributions of our study below:

% Particularly, the second smallest eigenfunction (also known as the Fiedler vector) can well cast the underlying affinities among the states, which helps to distinguish the states (locations) with low/high demands.
\begin{itemize}
    \item We introduce the deep relocation option policy (DROP) that resolves the exploration issue with decisive guidance for relocating vacant vehicles. DROP can be incorporated into existing model-free advances for further improvements in ride-hailing applications. 
    \item We propose to develop DROP by learning from the system-level relocation behavior of all vehicle agents. This is achieved by constructing the Laplacian approximation of the network for all relocation trips as an embedding of the system-level relocation policy.  
    \item The DROP is able to strike a balance between exploring new relocation directions and utilizing past experiences through the pseudo-reward as a mixture of task-agnostic and task-dependent reward signals. This setting is particular suited for ride-hailing systems. 
    \item We use a custom-built high-fidelity simulator with real-world taxi trip record data to comprehensively demonstrate the performance of DROP. 
    % \item We report the generality of our option policy {\bf Xinwu, unclear what you mean by generality and how are we showing that without demonstration?} to other RL frameworks. We train the DQN-based option generator based on the sampled trajectories, which is independent of fetching the actual reward in the environment. Therefore, it is feasible to additionally combine the option DQN with the primitive options, indicating the generality under tailored design and training process.
\end{itemize}

%The rest of the paper is organized as follows. The next section formulates the vehicle relocating as a semi-Markov decision process and introduces the basic elements and settings in detail. Section 3 proposes the architecture of our RL model under the option setting and introduces our high-fidelity simulator. Section 4 performs extensive numerical experiments and discusses the results. And the last section summarizes the study and highlights the main findings and future directions.
\section{Related work}
% \begin{itemize}
%     \item components in reward design
%     \item reward shaping details
%     \item model-based and model-free
% \end{itemize}

Existing solutions to the vehicle repositioning problem include model-based and model-free methods. The model-based approach is based on formulating the transition of the states by abstracting the dynamics of the underlying system~\cite{pavone2012robotic,smith2013rebalancing,zhang2016control,chuah2018optimal}. However, the intrinsic interactions among large-scale vehicle agents can be hardly characterized by aggregated closed-form equations. On the other hand, the model-free approaches have exhibited great merits in handling the intricacies in the large-scale system. Particularly, the RL-based techniques have been applied to solve the practical problems in the ride-hailing system, such as order dispatching~\cite{li2019efficient,qin2020ride} and  joint order dispatch and vehicle repositioning~\cite{zhou2019multi,holler2019deep,al2019deeppool,haliem2020distributed}, and most recently, option-based vehicle repositioning~\cite{tang2019deep,jiao2021realworld}. For the vehicle repositioning problems, we briefly summarize three key components for the framework design: (1) improvement on the policy-/value-based network~\cite{jin2019coride,holler2019deep,wang2020joint,jiao2021realworld}, (2) reward shaping~\cite{tang2019deep,wang2019adaptive,shou2020reward}, and (3) state representation design~\cite{al2019deeppool,tang2019deep,schmoll2020semi}. Interested readers are referred to~\cite{qin2021reinforcement} for a detailed review on RL in vehicle repositioning problem.

%  can dynamically learn the transition probability by interacting with the environment. The DRL-based approaches
% Specifically, Holler et al.~\cite{holler2019deep} report the advance of deep Q-network (DQN) based method of the sample efficiency. Tang et al.~\cite{tang2019deep} propose the tailored reward design under the options framework by uniformly distributing the total revenue over a wide-range time horizon. As such

% \begin{itemize}
%     \item requirement for lower sample complexity: previous work and recent work (find the one pre-defining cold-/hot-zones.)
%     \item the use of $\epsilon$-greedy and its limitation 
%     \item However, none of the existing work ...., mechanism on network level
% \end{itemize}

Sample complexity acts as one key challenge to enhance the process efficiency for a practical mechanism for vehicle repositioning~\cite{qin2021reinforcement}. To improve the sample complexity, one solution is to manually design the high-level policy using the domain knowledge of the ride-hailing system (e.g., pre-specifying the supply-demand distribution~\cite{xu2018large}). However, such handcrafted heuristics can hardly adapt to the varying supply-demand pattern in a large-scale system. Another solution relies on the option framework~\cite{brunskill2014pac,machado2017laplacian}, where the option supervises the agent to conduct decisive exploration on the environment, thereby escaping from bottleneck states of ``locally dithering'' (e.g., in low-demand or oversupplying regions).

So far, for real-world ride-hailing systems, there is no existing study that provides a general solution that resolves sample complexity issue that arises from exploration in model-free approaches. This highlights an important research direction, especially considering the scale of the system and the practical importance, which also motivates our study in exploring option policies as a potential remedy for improving exploration efficiency. 
\section{Preliminaries}
We formulate the ride-hailing vehicle repositioning problem as a semi-Markov Decision Process (SMDP~\cite{sutton1999between}) with $N$ vehicle agents. The relocation is based on the hexagonal cells $\mathcal{H}$ over discrete time steps $t \in \mathcal{T}:=\{0,1,2,\cdots,T\}$. 
% hexagonal grid system is commonly used to represent the spatial world for its same distance between neighbouring cell centers and good approximation of circles with an optimal perimeter/area ratio~\cite{hales2001honeycomb}.
Formally, we define the semi-Markov game as a tuple of six: $\mathcal{G} = \left(N,\mathcal{S},\mathcal{O},\mathcal{P},\mathcal{R},\gamma\right)$, where $N$ is the number of agents, $\mathcal{S}$ is the set of states, $\mathcal{O}$ is the set of options, $\mathcal{P}:\mathcal{S}\times\mathcal{O}\times\mathcal{S} \rightarrow [0,1]$ is the transition probability, $\mathcal{R}:\mathcal{S}\times\mathcal{O}\rightarrow\mathbb{R}$ is the reward function, and $\gamma$ is the discount factor. In this study, we define $\mathcal{O}=\mathcal{O_{DR}}\bigcup \mathcal{A}$. $\mathcal{A}$ is the set of temporally-extended primitive actions that an agent follow to relocate to one of the neighboring hexagon cells. $\mathcal{O_{DR}}$ is the deep relocation option policy (DROP) to be discussed in the rest of the paper, which contains a set of low-level policy $\omega_k$, each as an option for choosing a sequence of primitive actions. Detailed specifications of state, option, reward, and state transition in this study is summarized in the appendix. 

With the above definitions, we develop a model-free multi-agent reinforcement learning framework with options to solve the vehicle repositioning problem under the SMDP setting. Specifically, vehicle agents follow a high-level relocation policy $\pi_r$ to maximize their cumulative rewards as: 
\begin{dmath}
    Q_{\pi_r}(s_{t},o_{t})= r_{t+1}+\cdots+\gamma^{\Delta{t}-1} r_{t+\Delta{t}} + \gamma^{\Delta{t}} \max_{o_{t+\Delta{t}}} Q_{\pi_r}\left(s_{t+\Delta{t}},o_{t+\Delta{t}}\right),
    \label{eq:bellman_equaiton}
\end{dmath}
where $\Delta_t$ is the duration of the option $o_t$. The reward at each time tick $t$ incurs trip distance, time and passenger payments. Order dispatch is controlled by an exogenous policy $\pi_g$ (e.g., rule based or model based). In this study, we train $\pi_r$ using a single DQN shared by all vehicle agents, which allows for decentralized asynchronous action taking and centralized training with combined state transitions. %we follow the centralized-learning decentralized-execution framework where

\section{Deep relocation option policy}\label{sec:option_framework}
% This section introduces the development of DROP, including the construction of time-expanded relocation graph (TERG), the Laplacian approximation and the training algorithms. 
\subsubsection{Time-expanded relocation graph}
In a ride-hailing system, vehicle movements at each step can be categorized into occupied and unoccupied trips. The latter can be further divided into pick-up and relocation trips. The RL controller learns from transitions as the result of all vehicle trips. Nevertheless, we argue that occupied and pick-up trips are less informative than relocation trips in framing future relocation decisions. For instance, the number of trips between two locations carries no explicit knowledge on the demand-supply relationship at the two trip ends. On the other hand, the number of relocation trips at a location may serve as an important indicator of if the demand has been sufficiently served.

\begin{figure}
    \centering
    \includegraphics[width=1.0\linewidth]{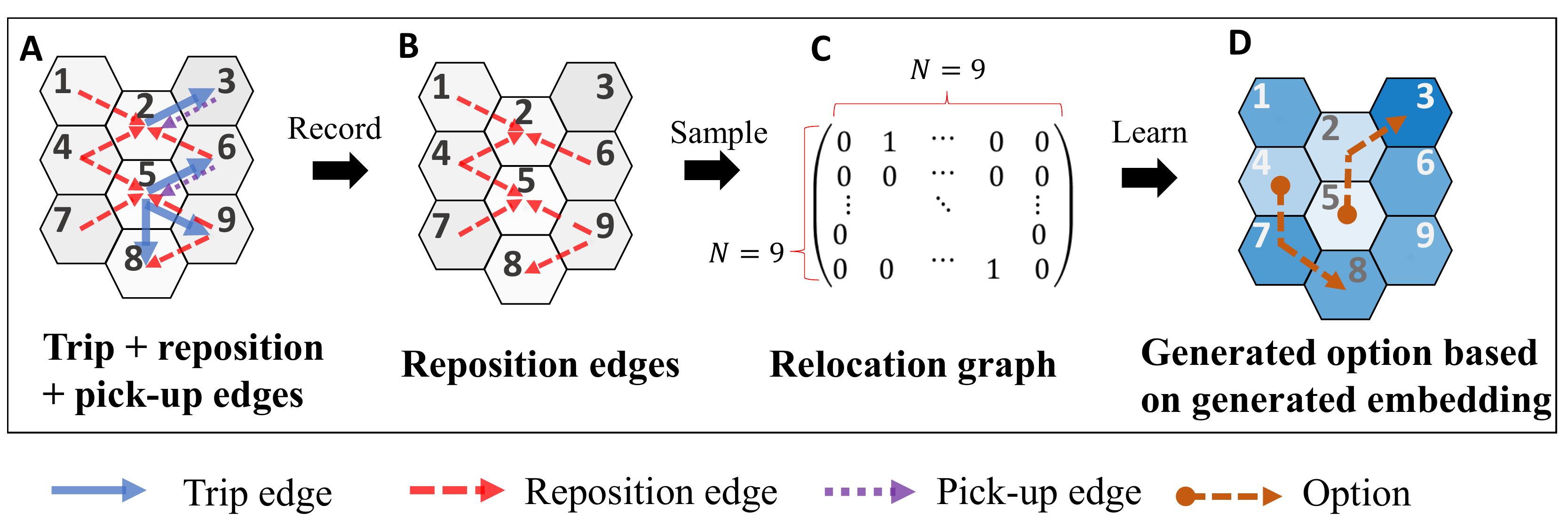}
    \caption{Flow chart of generating a relocation graph and the embedding representation}
    \label{fig:relocation_graph}
\end{figure}

We propose to construct DROP by mining additional information from transitions associated with relocation trips. This is achieved by constructing TERG $G_R(\mathcal{N},\mathcal{E})$ with its adjacency matrix denoted by $A_{G_R}$. For $G_R$, the set of nodes $\mathcal{N}$ corresponds to the set of states $s\in \mathcal{S}$ and we define
\begin{equation}
A_{G_R}(s,s')=\begin{cases}
n, \quad \text{if $s, s'$ are adjacent}\\
0, \quad \text{o.w.}
\end{cases}
\end{equation}
with $n$ denotes the number of relocation trips between the states $s$ and $s'$. The adjacency is satisfied if $s$ can reach to $s'$ via a primitive action,e.g., $s'$ is an immediate neighbor of $s$. We note that a relocation in the system is directional, nevertheless, we assume $G_R$ is undirected. This will not affect the interpretation since a relocation transition can only be observed if an agent completes an empty trip between the two states, implying lack of demand at both ends (also in the surrounding area). An example of TERG construction is shown in Figure~\ref{fig:relocation_graph} A-C.

As the relocation decisions are obtained from the RL model, $G_R$ naturally becomes a network representation of the policy $\pi_r(\theta_t)$ at the training step $t$. It captures how actions made by the system of agents are connected spatially and temporally. In this regard, a succinct embedding of TERG is deemed useful to framing a 'critic' for $\pi_r$.

% We here focus only on the set of relocation events as they carry the most important information that can contribute to differentiating the value of each intermediate state. 

\subsubsection{Laplacian approximation of TERG}
The sequence of node degrees in TERG provides a point understanding of the demand-supply relationship of a particular state $s$ following $\pi_r$. To construct decisive options that can guide effective relocation, we need to extend the point understanding into a contrastive relationship between each pair of nodes. Laplacian representation is one such approach to capture the geometry of the TERG by using eigenvectors of the Laplacian matrix as the state embedding. Formally, by considering $\textbf{L}$ as the Laplacian matrix, we can build a Laplacian embedding for TERG by solving:
\begin{align}
    \min_{\textbf{u}_1,\cdots,\textbf{u}_d} &\sum_{i=1}^D \textbf{u}_i^T \textbf{L} \textbf{u}_i \label{eq:eig_draw_obj}\\
    \text{s.t.}\quad &\textbf{u}_i^T\textbf{u}_j=\delta_{ij}, \forall {i,j}=1,\cdots,D
    \label{eq:eig_draw}
\end{align}
where $\textbf{u}_i\in \mathbb{R}^{1\times |S|}$ is the eigenvector corresponding to the $i_{th}$ smallest eigenvalue of $\textbf{L}$. $\delta_{ij}$ is the Kronecker delta which takes the value of 1 if $i=j$ and 0 otherwise.

Eqs~\ref{eq:eig_draw_obj} and~\ref{eq:eig_draw} follow from the spectral graph drawing problem~\cite{koren2003spectral}, with the $m_{th}$ entry in $\textbf{u}_1,\cdots,\textbf{u}_D$ used as the coordinate for the $m_{th}$ node to visualize the layout of a given adjacency matrix in the $D$-dimensional space. As for the TERG, the resulting embedding $\textbf{f}(s)=(\textbf{u}_1(s),\cdots,\textbf{u}_D(s))$ has significant implications in understanding both demand-supply relationship of a state and the affinity of two states. One of the key result can be seen as follows. 
\begin{prop}
$||\textbf{f}(s)||< ||\textbf{f}(s')||$ if state $s$ is adjacent to more relocation trips in the TERG than state $s'$.
\label{prop:position}
\end{prop}

The proof can be found in the appendix. With a Laplacian embedding $\textbf{f}(s)$ and the above property, we highlight below several key observations that characterizes the current system dynamics under $\pi_r$:

\begin{enumerate}[label=\textbf{\textsc{S\arabic*:}},topsep=-1ex,itemsep=-1ex,partopsep=1ex,parsep=1ex]
    \item A high value of $||\textbf{f}(s)||$ indicates that the state $s$ has been seldom reached through relocation trips than other states. 
    \item A low value of $||\textbf{f}(s)||$ implies the state $s$ has been frequently visited by relocation trips than other states. 
    \item For a pair of state $s,s'$, larger $||\textbf{f}(s)-\textbf{f}(s')||$ values (e.g., the euclidean distance in the embedding space) suggest that few relocation trips are made between the two states following $\pi_r$.
    \item For a pair of state $s,s'$, a small $||\textbf{f}(s)-\textbf{f}(s')||$ value indicates that the two states are frequently connected by relocation trips following $\pi_r$.
\end{enumerate}

%{\bf While above observations are good, there is nothing formal. Can something be proven for the relations. Also, origin of what?}

The above observations hint at immediate directions for relocation option construction that are task agnostic and can facilitate exploration. For instance, instead of random exploration, empty vehicles at the states with low $||\textbf{f}(s)||$ can be routed to a future state $s'$ that is distant from $\textbf{0}$. On the other hand, empty vehicles in states with high $||\textbf{f}(\cdot)||$ may imply the lack of demand at the associated time and location and can be positioned to locations with low $||\textbf{f}(\cdot)||$ values. However, we note that the above relocation directions are still ambiguous in the sense that a high or low $||\textbf{f}(\cdot)||$ does not warrant a rewarding state. The relocation option constructed is by no means "decisive" as a vehicle may move from a place with excessive supply to a place with no demand at all. We will address this issue in the next section. 

On the other hand, deriving the Laplacian embedding by Eq~\ref{eq:eig_draw_obj} is impractical for TERG, as $A_{G_R}$ is not tractable with the enormous size of the state space. A remedy of this issue is to construct Laplacian approximations using function approximators $\phi(s)=[f_1(s),\cdots,f_D(s)]$, which can be made possible with the deep neural network. We follow the recent work~\cite{wu2018laplacian} to derive the approximated Laplacian representation by converting the above graph drawing objective into minimizing the following unconstrained problem :
\begin{equation}
    G(f) = \frac{1}{2}\mathbb{E}_{s\sim\rho,s'\sim P^{\pi_r}(\cdot|s)} \sum_{i=1}^D [f_i(s)-f_i(s')]^2 + \lambda H(f)
    \label{eq:approximate_laplacian}
\end{equation}
\begin{equation}
    H(f) = \mathbb{E}_{s,s'\sim \rho} \sum_{i=1}^D\sum_{j=1}^D [f_i(s)f_j(s)-\delta_{ij}][f_i(s')f_j(s')-\delta_{ij}]
    \label{eq:repulse}
\end{equation}
The function $G(\textbf{f})$ is a combination of an attractive term for $\textbf{f}(s),\textbf{f}(s')$ and a repulsive term $H(\textbf{f})$ with the Lagrangian multiplier $\lambda$ enforcing the orthonormal constraint in Eq~\ref{eq:eig_draw}. By minimizing $G(\textbf{f})$, the attractive term will place states with more relocation trips in between closer and the repulsive term eliminates trivial solutions like $f_i(s)=0$. Readers are referred to~\cite{wu2018laplacian} for detailed derivation of the Laplacian approximation.

\begin{algorithm}[ht]
\caption{Deep relocating option generation}
\label{alg:drop_algo}
\textbf{Input}:exploitation coefficient $\alpha$, number of options $K$, option warm-up steps $N$, episode id $e$, sample transition trajectories at time of generation $\mathcal{T}^e$ \\
% \textbf{Parameter}: Optional list of parameters\\
\textbf{Output}: The set of trained DROP $\mathcal{O}_{DR}^e=\{\omega_{1}^e,\omega_{2}^e,..,\omega_{K}^e\}$.
\begin{algorithmic}[1] %[1] enables line numbers
\STATE Train eigen-representation of the TERG $\phi^e$ by minimizing $G(f)$ in Eq.~\ref{eq:approximate_laplacian} using samples from $\mathcal{T}^e$. 
\STATE Set $\mathcal{O}_{DR}^t\leftarrow \emptyset$
\FOR {$k=1,\cdots,K$}
\FOR {$n=1,\cdots,N$}
\STATE Train low-level policy $\omega_k^e(\phi^e)$ with reward $r_k$ (see Eqs.~\ref{eq:long_dist1}-\ref{eq:large_diff}).using samples from $\mathcal{T}^e$ %{\color{red}\bf where is low-level policy described? $n$ is not used here?}
\ENDFOR
\STATE $\mathcal{O}_{DR}^e\leftarrow \mathcal{O}_{DR}^e\bigcup \omega_k^e$
\ENDFOR
\STATE \textbf{return} $\mathcal{O}_{DR}^e$
\end{algorithmic}
\end{algorithm}

\subsubsection{Learning deep relocation policy}
The Laplacian representation for TERG provides the contrastive relationship between pairs of states that will allow for the construction of DROPs. This task-agnostic option construction approach is commonly seen in goal-seeking problems with a stationary environment, e.g., the Gridworld and MujoCo environments. Unfortunately, the same approach will fail catastrophically in the ride-hailing setting since the environment we are learning from is a network realization of the relocation policy $\pi_r$ that is dynamically changing. A rewarding option at one episode will likely become the source of oversupply in the next episode when multiple vehicle agents from other states exploit a target state. 

To account for the varying environment dynamics after learning $\phi$, we propose to use task-related signals $\mathbb{I}_{s'}$ which is an indicator variable: 1 if relocating to state $s'$ leads to a trip assignment following $\pi_d$ and 0 otherwise. By incorporating the trip signal into the pseudo-reward, the expectation of $\mathbb{I}_{s'}$ immediately captures the competition for a certain demand level. This also resolves the "non-decisive" issue mentioned above by only using task-agnostic signals from TERG. As a consequence, for each learned $\phi^i$, we train DROP by maximizing the following three pseudo-reward functions:
\begin{align}
    r_1^i(s,s')&=||\textbf{f}(s)||-||\textbf{f}(s')||+\alpha \mathbb{I}_{s'} \label{eq:long_dist1}\\
    r_2^i(s,s')&=||\textbf{f}(s)'||-||\textbf{f}(s)||+\alpha \mathbb{I}_{s'}  \label{eq:long_dist2}\\
    r_3^i(s,s')&=||\textbf{f}(s)-\textbf{f}(s)'||+\alpha \mathbb{I}_{s'} \label{eq:large_diff}
\end{align}
Specifically, Eqs.~\ref{eq:long_dist1} and~\ref{eq:long_dist2} correspond to S1 and S2 above, and the maximization of the accumulated rewards will enable vehicle agents to traverse to the target state with the minimum number of transitions. Eq.~\ref{eq:large_diff} is related to S3 and will motivate vehicle agents to explore neighbor states that are less-visited before. On the other hand, $\mathbb{I}_{s'}$ helps balance between exploration and exploitation. In the early stage of training, $\mathbb{I}_{s'}$ will be primarily zero, and DROP will follow signals from the Laplacian embedding. With more trip assignments identified after initial exploration, DROP will start exploiting routes with a higher overall chance of trips. We use $\alpha$ to control the trade-off between the two reward signals. $\alpha=0$ suggests pure exploration based on Laplacian representation, and $\alpha\rightarrow \infty$ only exploits trip signals explored so far. As a final remark, we note this mixture of rewards is well suited for an environment with multiple goal states, such as in ride-hailing systems. Unlike a system with a single goal state, task-dependent signals may also facilitate the learning process as the goal state is likely associated with a sequence of future rewards.

% With the learned Laplacian representation, we measure the distance of a state $s$ to $\textbf{0}$ as $d_{s}^i=||\phi(s)^i||$, and the distance between two states $s,s'$ as calculated as $d_{ss'}=||\phi(s)-\phi(s')||$. We then construct the following 3 option policy for each learned Laplacian representation with the designed Q function as:
% \begin{align}
%     &Q_{dr_1}(s,a)&=d_{ss'}(1+\alpha \mathbb{I}_{s'})+\gamma \max_{a'\in\mathcal{A}} Q_{dr_1}(s',a') \label{eq:long_dist1}\\
%     &Q_{dr_2}(s,a)&=(d_{s0}-d_{s'0})(1+\alpha \mathbb{I}_{s'})+\gamma \max_{a'\in\mathcal{A}} Q_{dr_2}(s',a') \label{eq:long_dist2}\\
%     &Q_{dr_3}(s,a)&=(d_{s'0}-d_{s0})(1+\alpha \mathbb{I}_{s'})+\gamma \max_{a'\in\mathcal{A}} Q_{dr_3}(s',a') 
% \end{align}
% where $\mathbb{I}_s'$ Moreover, we do not train deep relocation option policy under the SMDP setting since the temporal dimension has been implicitly accounted for while learning for the Laplacian representation. 
\subsubsection{Model training} We note that each DROP $\omega_k$ is trained under the standard MDP setting instead of SMDP following Eq.~\ref{eq:q_fun} since the temporal dimension has been implicitly accounted for while learning for the Laplacian representation. 
\begin{equation}
Q_{\omega_k}(s,a;\phi)=r_{\omega_k}+\gamma \max_{a'\in\mathcal{A}} Q_{\omega_k}(s',a';\phi), \forall k=1,2,3 \label{eq:q_fun}
\end{equation}
with $r_{\omega_k}$ corresponding to the above three rewards.

On the other hand, TERG may change rapidly as we train $\pi_r$. To address this issue, one may learn $\pi_r$ and the Laplacian representation simultaneously. But this may instead harm the performance of $\pi_r$ due to unstable options, as will be shown in the results. To overcome this issue, we propose to periodically (e.g., every 2-3 episodes) augment option $\mathcal{O}$ with a new set of DROPs $\omega_k^e(\phi^e)$ based on the fixed (like a target network) Laplacian representation $\bar{\phi}^e$ , which is trained using sample relocation transitions from the current $\pi_r(\theta^e)$ at episode $e$. Moreover, we still train each added DROP $\omega_k^e$ to cope with the latest trip reward signals following updated $\pi_r$, so that the added DROP is always relevant to the environment. We use DQN to train both Laplacian representation and the deep relocation option policy. We use a dedicated memory buffer $B$ to store relocation transitions. The algorithm for training DROP is summarized in Algorithm~\ref{alg:drop_algo}, and the algorithm for training high-level relocation policy $\pi_r$ of ride-hailing system is summarized in Algorithm~\ref{alg:overall_train}.

% The final issue to be addressed is the non-stationary issue that will arise in learning the TERG representation in a multi-agent environment. 

\begin{algorithm}[ht]
\caption{High-level relocation policy training}
\label{alg:overall_train}
\textbf{Input}: replay memory $M$, relocation state-transition memory $B$,  coefficient $\epsilon$  
\begin{algorithmic}[1] %[1] enables line numbers
\FOR {episode=1:Episodes}
\STATE Simulator initialization.
\FOR {t=1:T}
\STATE Update passengers and vehicles.
\STATE Record all state transitions in $M$.
\STATE Record relocation transitions in $B$. 
\STATE Get global state $s_t^g$ and local state $s_{i,t}^l$ for all idle vehicles
\STATE For each idle vehicle $i$, choose random option $o_i^t$ from $\mathcal{O}$ with probability $\epsilon$; otherwise obtain relocation option $o_i^t\leftarrow\argmax_o Q_{\pi_r}(s_g^t,s_i^t,o;\theta)$. 
\IF {$o_i \in \mathcal{O}_{DR}^t$}
\STATE $a_i^t\leftarrow \argmax_a Q_{w_{o_i}}(s_g^t,s_i^t,a;\theta_{w_{o_i}})$
\ELSE
\STATE $a_i^t\leftarrow o_i^t$
\ENDIF
\STATE Train $\pi_r$ with samples from $M$.
\STATE Train each $\omega_k$ with samples from $B$. 
\ENDFOR
\STATE For every $p$ episodes, generate new set of relocation options $\mathcal{O}_{DR}^e$ following Algorithm~\ref{alg:drop_algo} using samples from $B$. Set $\mathcal{O}\leftarrow\mathcal{O}\bigcup\mathcal{O}_{DR}^e$.
\ENDFOR
\end{algorithmic}
\end{algorithm}

\begin{table*}[ht]
\centering
\caption{Perform comparison with 7 days of weekday trip data. }
\label{tab:validation}
\begin{tabular}{l|llll|llll}
\toprule
\multirow{2}{*}{Model/Metrics} & \multicolumn{4}{c}{Hourly revenue per vehicle}       & \multicolumn{4}{c}{Average rejection rate} \\
                               & Peak & Off-peak & Night & Overall & Peak  & Off-peak  & Night & Overall \\
                               \midrule
DQN                            & 17.7 $\pm$ 1.9  & 18.7 $\pm$ 1.1   & 4.6 $\pm$ 0.9     & 15.3 $\pm$ 1.0     & 22.3   & 8.3    & 0.9   & 17.3     \\
DRDQN                        & \textbf{22.7 $\pm$ 0.5}  & \textbf{20.9 $\pm$ 1.1}   & \textbf{5.2 $\pm$ 0.9}     & \textbf{17.7 $\pm$ 0.6}     & \textbf{4.1}   & \textbf{2.4}    & \textbf{0.11}    & \textbf{7.3}     \\
% DRDQN-1                        & 21.1 $\pm$ 0.6  & 20.5 $\pm$ 0.9   & 5.0 $\pm$ 0.8     & 17.1 $\pm$ 0.6     & 10.2   & 3.4    &  \textbf{0.04}     & 10.2     \\
DRDQN-inf                      & 20.7 $\pm$ 0.6  & 20.5 $\pm$ 0.9   & 5.0 $\pm$ 0.8     & 16.9 $\pm$ 0.7     & 11.8   & 3.7    & 0.2     & 10.9     \\
DRDQN-0                      & 16.7 $\pm$ 2.4  & 18.4 $\pm$ 1.4   & 4.4 $\pm$ 0.8     & 14.5 $\pm$ 1.6     & 26.9   & 10.1    & 3.4     & 21.5     \\
ODRDQN                       & 20.3 $\pm$ 0.7  & 19.8 $\pm$ 0.9   & 5.0 $\pm$ 0.9 & 16.5 $\pm$ 0.7     & 12.9   & 4.9    & 0.2     & 12.0     \\
RDRDQN                       & 17.5 $\pm$ 2.8  & 18.8 $\pm$ 1.6   & 4.6 $\pm$ 1.0     & 14.9 $\pm$ 1.9     & 23.8   & 8.7    & 4.0     & 20.1     \\
\hline
Greedy                         & 18.2 $\pm$ 1.9 & 20.6 $\pm$ 0.8   & 4.6 $\pm$ 0.8     & 14.9$^*$ $\pm$ 0.4    & 7.3   & 0.2 & 0     & 8.4     \\
Random                         & 12.6 $\pm$ 0.3  & 16.0 $\pm$ 0.8   & 3.6 $\pm$ 0.5     & 11.3 $\pm$ 0.5     & 45.2   & 19.4    & 15.7     & 37.8    \\
\bottomrule
\end{tabular}

\raggedright Peak: 8-10 AM and 6-8 PM / Off-peak: 10 AM-2 PM / Night: 3-5 AM

*: Reward of greedy is achieved by assuming unlimited matching radius between drivers and vehicles.
\end{table*}

\section{Experiments}
% In this section we present the simulation environment and real-world data used for training the algorithm, and report comprehensive empirical evaluations of the proposed deep relocation policies. 

\subsubsection{Study area and environment setting}
We demonstrate the effectiveness of DROP with a case study of the taxi system in New York City (NYC). In particular, we developed a high-fidelity simulator (see Appendix B for details) and used NYC taxi data during May 2016 as the input~\cite{nyc_taxidata2016}. We consider each simulation tick as 1 minute, and each episode has 1440 ticks. The study area is divided into $1347$ hexagon cells (each covering an area of 0.14 $mi^2$), and we assume all trips are generated and served at the centroid. The travel time between each pair of centroids is obtained by overlaying with the actual road network and then querying from the OSRM engine. 

We conduct a down-scaled experiment of the taxi system with 500 vehicle agents, and we obtain the demand profile by sampling 10\% of the historical trip record uniformly at random (around 40k trips). Unless otherwise specified, available vehicles will be dispatched to passengers that can be reached within 8 minutes, and the maximum waiting time for passengers without vehicle assignments is 5 minutes. At the beginning of each simulation, the vehicle agents will enter the market following a uniform distribution across the first 30 minutes and over the study area. The first 14 weekdays are used for model training and the following 7 weekdays for model validation. More detailed specifications on modeling parameters and network structure can be found in Appendix D.

\subsubsection{Models}
To validate the effectiveness of DROP, we conduct comparison among the following models
\begin{enumerate}
    \item \textbf{DRDQN} represents the deep relocation DQN model, which is the proposed approach with $\alpha=1$ for the mixture of pseudo-reward.
    \item \textbf{RDRDQN} refers to DRDQN with random pseudo-reward, where we replace the values of Laplacian approximation with a value generated uniformly at random between 0 and 1 during the training process.
    \item \textbf{ODRDQN} refers to DRDQN with online training of DROP. In this case, we train DROP along with DRDQN and we do not expand $\mathcal{O}_{DR}$ over time.  
    \item \textbf{DRDQN-0} is the DRDQN with $\alpha=0$. The training of DROP is therefore task agnostic. \item \textbf{DRDQN-inf} is the DRDQN with $\alpha\rightarrow\infty$. The training of DROP completely depends on trip signals. 
    \item \textbf{DQN} represents the baseline model without DROP. 
    \item \textbf{Random:} The relocation is performed by selecting from primitive actions uniformly at random.
    \item \textbf{Greedy} in this study refers to a global matching policy that pairs vehicle agents and trip requests without maximum matching distance constraint. In this case, no relocation is performed if demand is greater than supply, otherwise it will perform Random policy. 
\end{enumerate}

\begin{figure*}
    \centering %0.23， 0.75
    \subfloat[Distribution of requests.]{\includegraphics[width=.2\linewidth]{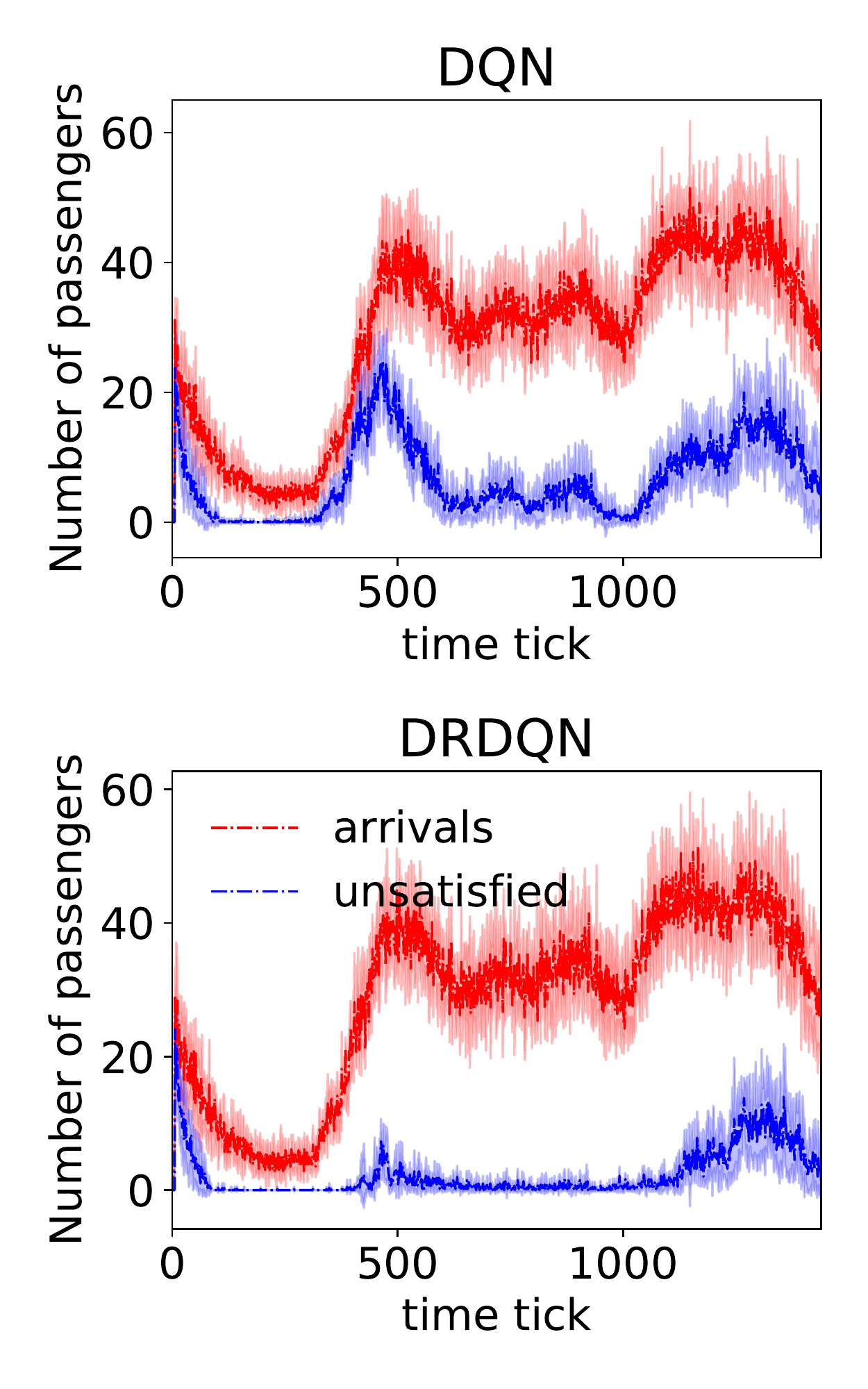}\label{fig:serve_leave}}
    \subfloat[Spatial distribution of demand-supply gap]{\includegraphics[width=.7\linewidth]{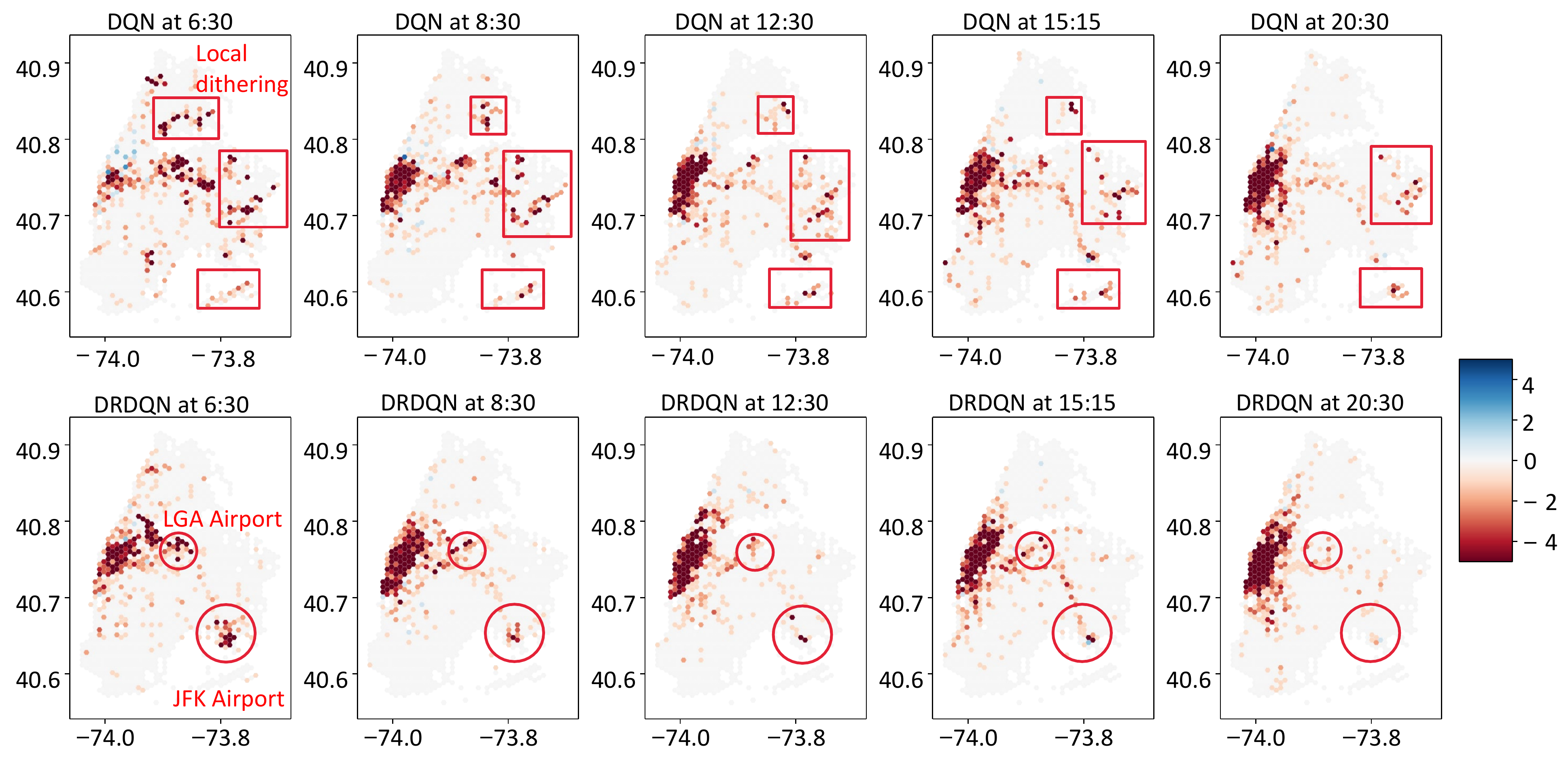} \label{fig:ds_gap}}
    \caption{System dynamics following DQN and DRDQN. }
\end{figure*}

\subsection{Results}
\begin{figure}[ht]
    \centering
    \includegraphics[width=.8\linewidth]{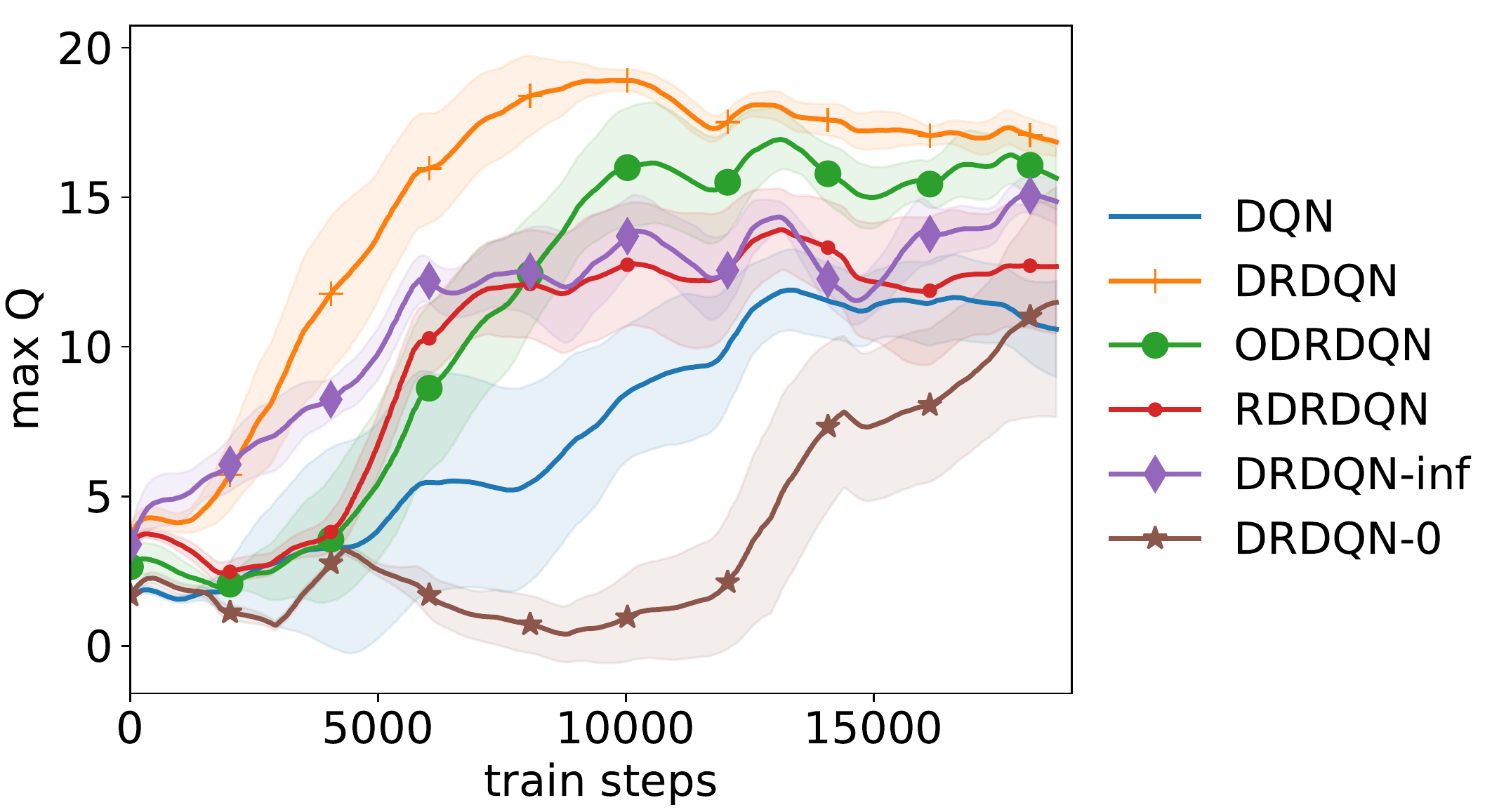}
    \caption{Convergence of maxQ for training. Results are the moving average of 720 training steps and the shaded area represents $\pm$ 0.5 std. }
    \label{fig:training}
\end{figure}

The results of model training are summarized in Fig.~\ref{fig:training} where maxQ values are visualized along with a 0.5 standard deviation. It can be found that all maxQ values are converged before the training concludes. Except for DRDQN-0, all other variations of DRDQN are shown to outperform the vanilla DQN model at the training stage. The training results of DRDQN-inf highlight the importance of task-related reward signals for developing option policies for ride-hailing systems. Most importantly, though DRDQN-0 may fail on its own, we report that DRDQN with a combination of task-agnostic and task-dependent pseudo-rewards will significantly improve the model performance. Note that DRDQN-inf can reach a high maxQ value in the early stage by exploiting trip signals. DRDQN further this gain with a balance of exploring new areas and reinforcing past achievements through the help of DROP. 

The superior performance of DROP can be further justified by inspecting the validation results as shown in Tab.~\ref{tab:validation}. The table summarizes the average hourly revenue per vehicle, along with the standard deviations, during different time periods. The average rejection rate is also reported as the ratio between unsatisfied passengers and all passenger arrivals. We observe that the vanilla DQN may be considered a good controller when compared with the Random and Greedy policies. Its improvement over Greedy policy also asserts the importance of vehicle relocation in ride-hailing systems. Note that the Greedy policy achieves the lowest rejection rate by violating the maximum waiting time constraint, and we use the results mainly to demonstrate that the case study is of balanced fleet size and demand level when optimized. On the other hand, we report that DRDQN achieves the best performance as compared to all other benchmark models. The incorporation of DROP enables a significant improvement over the vanilla DQN model, both in terms of average performance (a 15.7\% improvement over the entire episode) as well as more consistent performance across different random seeds. The biggest gain results from the peak hour period, where DRDQN can satisfy 95.9\% of all requests and obtain 28.2\% more hourly revenue than DQN. The ODRDQN is also found to outperform DQN but being inferior to the DRDQN model. This highlights the importance of avoiding frequent updates of embedding for TERG. Instead, inserting new options with fixed representation every few episodes will help stabilize the training process. 

\subsubsection{Service performance} Fig.~\ref{fig:serve_leave} shows the per-tick arrival of passenger requests and the unsatisfied number of passenger requests. At the beginning of each episode, there are many unsatisfied requests since vehicles uniformly enter the system within the first 30 ticks. Both models well serve the nighttime period with a low demand rate, and only DRDQN can carry superior performance over the daytime. However, neither models is able to meet the excessive demand late in the night with the given fleet size. 

The reasons for DROP being superior can be well interpreted by visualizing the demand-supply gap at different times of the day, as shown in Fig.~\ref{fig:ds_gap}. The results demonstrate the distribution of all vehicle agents, regardless of occupied or unoccupied, and a negative value indicates more vehicle agents than the demand rate at the location. The figure presents an appealing observation of how well DROP addresses the dithering issue in the sparse reward problem. We see that vanilla DQN consistently suffers from local dithering throughout the day, primarily in the peripheral areas with few requests. On the other hand, such an issue is completely resolved by the DRDQN model. More importantly, the DROP also learns to align vacant vehicles perfectly with high-demand areas such as Manhattan, JFK, and LGA airports. It also learns to pre-position vehicles into the airport (see DRDQN at 6:30) to prepare for the arrival of requests from early morning flights. 
\begin{figure}
    \centering
    \includegraphics[width=\linewidth]{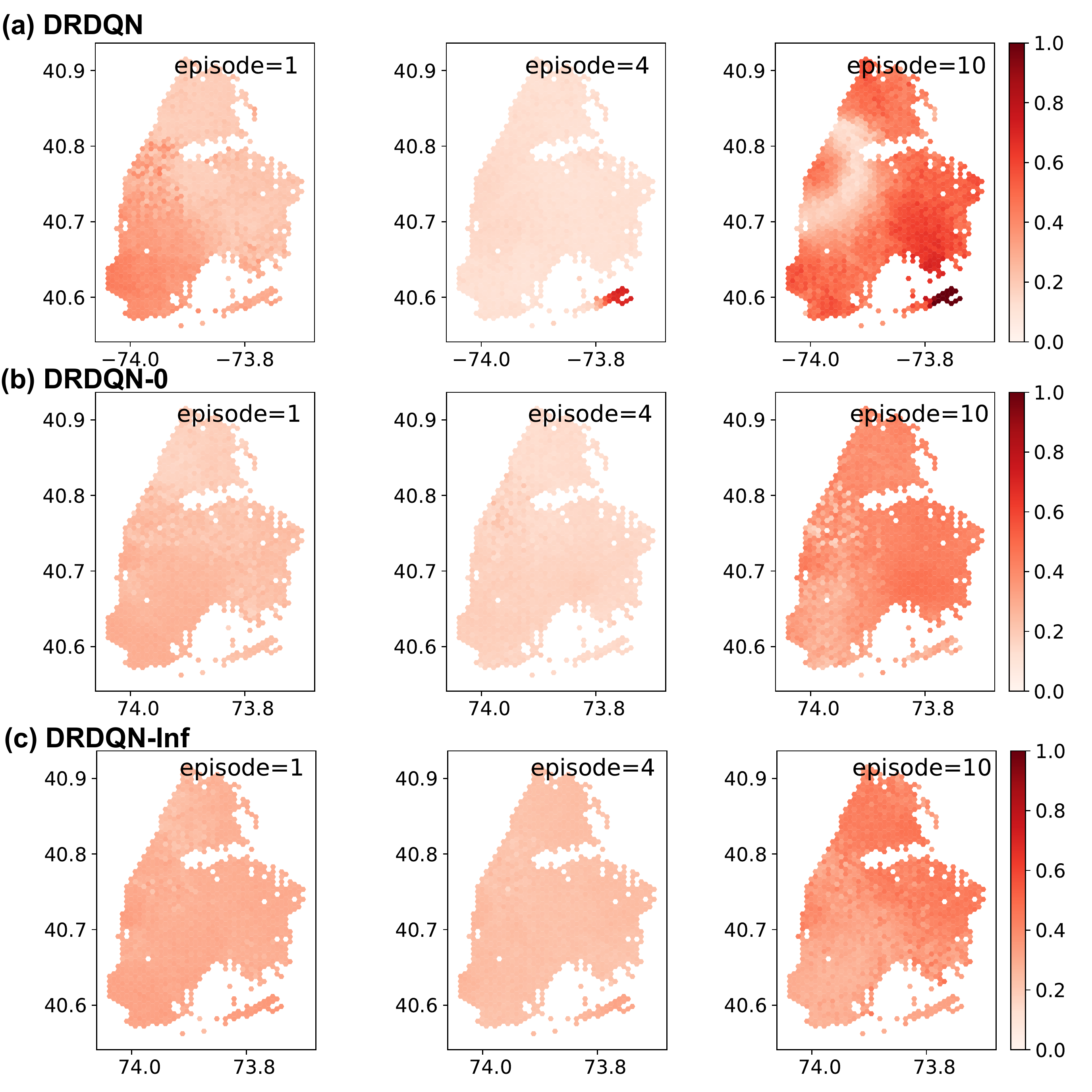}
    \caption{Example of learned Laplacian embedding $||\textbf{f}(s)||$ at 3 PM for different episodes.}
    \label{fig:embedding_overall}
\end{figure}
\begin{figure}
    \centering
    \includegraphics[width=1\linewidth]{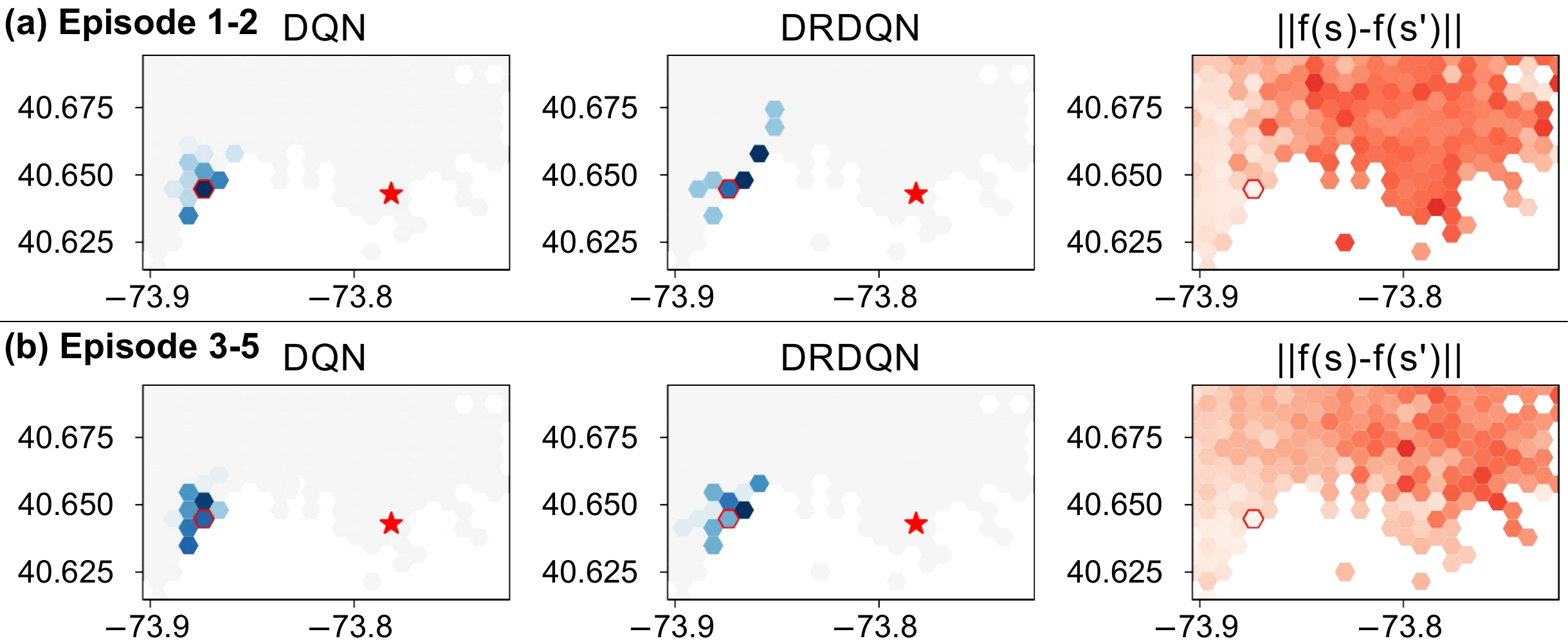}
    \caption{Learned policy over episodes between DQN and DRDQN (red star here refers to the JFK airport)}
    \label{fig:embedding_detail}
\end{figure}
\subsubsection{Learned representations} A key to the success of DROP is the learned representation of the TERG. Here we further examine how the Laplacian approximation of TERG helps DROP to identify better relocation policies. We first show in Fig.~\ref{fig:embedding_overall} samples of learned Laplacian embedding at 3 PM from added DROPs by the end of episodes 1, 4, and 10. Recall the norm $||f(s)||$ describes the supply level, and we observe all three models learn similar representations at the beginning: there are more relocation trips in the upper-right half of the study area than the lower-left part. Inspired by the learned representations, the first set of DROPs are added and followed by agents that lead to the updated observation by the end of episode 4. The incorporation of DROP leads to more balanced relocation dynamics amid the learning process. This finding is largely consistent across all models, with the only exception that the DRDQN already learns not to relocate to the lower right corner. With the further addition of DROPs, we observe different learned representations by the end of episode 10. For DRDQN-0 and DRDQN-inf, we note the oversupply pattern is fully reverted as compared to episode 1, with more relocation trips in the lower-left half of the study area. This is the result of excessive exploitation (DRDQN-inf) and exploration (DRDQN-0). On the other hand, we report that the system following DRDQN arrives at an ideal relocation dynamics: there are few relocation trips in both high-demand and low-demand areas (darker color), and a balanced demand-supply pattern is realized. Moreover, the relocation dynamics construct a 'halo' in between the highest-demand area (Manhattan) and the remaining boroughs of NYC. In this case, a vehicle agent can easily reach either side of the halo and quickly reach future requests at a low cost.

Finally, we show in Fig.~\ref{fig:embedding_detail} how the learned representation contributes to the formation of effective DROP. The figure compares how vehicle relocates starting from the hexagon cell with red edges. We also mark by the star the JFK airport as a major trip generation point nearby. Hexagon cells with a darker color are associated with more relocation trips. Consistent with previous discussions, we observe severe dithering issue for the DQN model. Vehicles are unable to escape the neighboring areas and keep coming back. As a result, the DQN learns to stay at the same location due to the lack of reward signals. On the other hand, DRDQN first constructs relocation options heading to neighboring areas with high $||\textbf{f}(s)-\textbf{f}(s')||$ values within the first two episodes. It subsequently learns a relocation policy pointing into the direction of the trip reward signal at the JFK airport within episodes 3-5. 
% Please add the following required packages to your document preamble

\section{Conclusion}

We study the optimal vehicle relocation problem for ride-hailing systems and we propose the deep relocation option policy (DROP) to address the exploration issue that will arise in the environment with sparse reward and unbalanced demand and supply levels. We introduce the algorithms for training DROPs and augmenting the high-level relocation policy. The results show that DROP can improve base line DQN by over 15.7\% and also achieves the lowest rejection rate. The learned embeddings are also visualized to support the reasonings behind DROP development. Future work will extend to the group-based DROP framework which may allow agents at the same location to perform different options. 

% \newpage
\bibliographystyle{aaai22.bst}
\bibliography{aaai22}
% \newpage
\appendix
\onecolumn
 \setcounter{secnumdepth}{1}
\section*{Supplementary Material for 
\\``DROP: Deep relocating option policy for optimal ride-hailing vehicle repositioning"}
%\section{Appendix}

\section{Modeling preliminaries}

\subsubsection{State} We consider both global state $\mathcal{S}^{g}$ and local state $\mathcal{S}^{l}$. At time step $t$, all agents receive the same global state (e.g., via by a central controller), denoted by $s_{t}^{g}\in \mathcal{S}^{g}$, including supply-demand patterns at time $t$. The local state for agent $i$  at time $t$, $s_{i,t}^{l}\in \mathcal{S}^{l}$, can be expressed as a tuple $\left(t,\mathbf{m}_{i,t}\right)$, with $\mathbf{m}_{i,t}$ an $N$-dimension one-hot vector of $i$'s location. 
% We additionally append a static hexagonal-level diffusion of dimension $|\mathcal{M}|$, with each element being the summation of diffusion probability, denoted by the radial basis function of trip time between two hexagon grid pairs. 
% , and $\mu_{i,t}$ represents the status of agent $i$, given as  $\mu_{i,t}=0$ if agent $i$ is cruising and $\mu_{i,t}=1$ if agent $i$ is serving passenger.
% The dimension of state space is $|\mathcal{M}| \times 2 + 3$.

\subsubsection{Option} The option is a set of temporally-extended actions in the context of SMDP~\cite{sutton1999between}, which can well describe the movement of vehicle agents with different durations (e.g., relocating, picking up or serving a passenger).
The option can be formally expressed as a tuple: $(\mathcal{I}, \pi, \zeta)$, where $\mathcal{I} \subseteq \mathcal{S}$ is the set of states where the option can initiate, $\pi: \mathcal{S} \times \mathcal{O} \rightarrow [0,1]$ is the option policy, and $\zeta: \mathcal{S} \rightarrow [0,1]$ is the termination condition, quantified by the probability of encountering the terminal states. 

In this study, we consider $\mathcal{O}=\mathcal{O_{DR}}\bigcup \mathcal{A}$. $\mathcal{A}$ is the set of primitive actions that allow vehicle agents to relocate to one of the nearest six neighbors or stay at the current hexagon. The corresponding terminating condition for each primitive action is therefore the arrival at the destination. On the other hand, $\mathcal{O}_{DR}$ is the set of deep relocation option which will be explained in later sections. Each $o_k^{dr}\in \mathcal{O}^{DR}$ is associated with a low level policy $\omega_k: \mathcal{S}\times\mathcal{A}\rightarrow[0,1]$ that will perform a sequence of primitive actions. In this study, we set the option $o_k^{dr}$  to last a fixed $L$ steps so that the termination condition is well-defined.

% Specifically, the terminal condition $\beta$ is triggered by vehicle arriving at the option-targeted hexagon or being matched while repositioning. Further details will be introduced in section~\ref{sec:option_framework} {\bf Xinwu, we cannot have section numbers here - can do for Appendix though. }. 

% ($m_{t} = m_{o}$ and $m_{t-1} \ne m_{t}$) ;  ($\mu_{t-1}=0$ and $\mu_{t}=1$) 

% At every time step, the platform assigns options to the idle vehicles either relocate to one of the six nearest hexagons or stay at the current hexagon. In this regard, the option space $\mathcal{O}$ is of $7$ dimensions.

% Specifically, the statuses of an agent consist of idled, assigned, occupied, and cruising. 
\subsubsection{State transition} The controller assigns relocation options to idle vehicles. Denote the time gaps between two consecutive decision-making processes by transition ticks. We consider a transition starting at time $t$ and last for $\Delta{t}$. Following~\cite{jiao2021realworld}, we explicitly define the overall policy as: $\pi = \pi_{r} \bigcup \pi_{d}$, where $\pi_{r}$ and $\pi_{d}$ represent the \textit{reposition} and \textit{dispatching} policies. At time step $t$, the idle vehicles first execute the relocating options following $\pi_{r}$. Next, the active (i.e., cruising and idling) vehicles are assigned with open orders under an exogenous dispatching policy $\pi_{d}$. Finally, the relocating option based on $\pi_{r}$ will guide the agent to a new state at the next transition tick $t+\Delta t$. Note that one transition starting at $t$ only contains one serving or relocating cycle, such that the incurred reward $r_{t}$ only depends on the initial option $o_{t}$. Any further decision-making processes will trigger new transitions (e.g., being matched during the relocating cycle), thus the new relocating option is independent of its preceding reward.

\subsubsection{Reward} At each time step $t$, the vehicle agent $i$ receives a reward $r_{i,t}$ that reflect the cost of travel distance and time and the payment from a passenger $j$ if the trip ends at $t$:
% The cumulative discounted return $R_{i}$ can be expressed as below:

% \begin{equation}
%     R_{i} = E\left[\sum_{\tau=1}^{\Gamma_{i}} \gamma^{\tau} r_{i,\tau}\right]
% \end{equation}
% where $\gamma \in [0,1]$ is the discount factor, $\Gamma_{i}$ is the total number of transitions for agent $i$ in one episode, $r_{i,\tau}$ is the reward received by agent $i$ between transition tick $\tau$ and $\tau+1$.

% We consider the reward between two transition tick $\tau$ and $\tau+1$ as the cumulative reward received from time step $t_{\tau}$ to $t_{\tau}+k_{\tau}$, where $k_{\tau}\ge 1$ is the elapsed time ticks during the transition $[\tau,\tau+1)$. The reward for agent $i$ at time step $t \in [t_{\tau},t_{\tau}+k_{\tau})$ can be expressed as below:

\begin{equation}
    r_{i,t} = \beta_{1} c_{i,t} - \beta_{2} \Delta{\tau}_{i,t} - \beta_{3} \Delta{d_{i,t}}
    \label{eq:reward}
\end{equation}
where $\beta_{1}, \beta_{2}, \beta_{3}$ are the weights for order payment, time cost, and distance cost, $c_{i,t}$ is the payment, $d_{i,t}$ is the travel distance, and $\Delta{\tau}_{i,t}$ is the length of the elapsed time ticks. For $c_{i,t}$, it includes the distance and time based fare minus a disutility term that accounts for the passenger waiting time. 

We adopt a time discounting technique as in~\cite{tang2019deep} to compute the cumulative rewards within transition tick $[t,t+\Delta t)$, and the term of cumulative reward in Eq.~\ref{eq:bellman_equaiton} can be rewritten as:

\begin{dmath}
    r_{i,t}(s_{t},o_{t}) = r_{i,t+1}+\cdots+\gamma^{\Delta{t}-1} r_{i,t+\Delta{t}}
    =\frac{r_{i,t}}{\Delta{t}}+ \gamma  \frac{r_{i,t}}{\Delta{t}}+\cdots+\gamma^{\Delta{t}-1} \frac{r_{i,t}}{\Delta{t}} = \frac{r_{i,t}\left(\gamma^{\Delta{t}}-1\right)}{\Delta{t}\left(\gamma-1\right)}
    \label{eq:decayed_reward}
\end{dmath}
where $r_{i,t}(s_{t},o_{t})$ is the discounted accumulative reward by taking option $o_{t}$ at state $s_{t}$ in the transition from $t$ to $t+\Delta t$, and $r_{i,t}$ is the total reward within $[t,t+\Delta{t})$ for agent $i$.

% For agent $i$, $c_{i,t}$ is the order payment at $t+k$ which is uniformly distributed along the $k$ time steps between $\tau$ and $\tau+1$, $\Delta{t_{i,t}}$ and $\Delta{d_{i,t}}$ are the elapsed time and distance between time step $t$ and $t+1$. We then apply a discount factor to approximately {\bf TO COMPLETE}

\section{Simulator}
To train and evaluate our model, we develop an agent-based simulator to characterize the interactions between agents (vehicles) and the environment (requests) using real-world taxi trip data~\cite{nyc_taxidata2016}. Enabled by Open Source Routing Machine (OSRM) engine~\cite{luxen-vetter-2011}, we can update the location of the vehicle agents at every tick and trace the their real-time dynamics (e.g., trip mileage, time to destination, etc.), which can further measure the rewards. Other basic settings for the simulator can be found in~\cite{haliem2020distributed}. We summarize main simulator components below. 

% \begin{figure}[!htbp]
%     \centering
%     \includegraphics[width = 0.5\linewidth]{figure/sim_flow_chart.pdf}
%     \caption{Simulation flow Chart}
%     \label{fig:sim_flow_chart}
% \end{figure}

%  get global state & local state -> attach actions that are generated from DQN-> passenger arrive / update -> vehicle dispatch -> vehicle match -> vehicle update -> vehicle step

\noindent \textbf{Vehicle status updates:} The vehicle statues are updated per tick with a new position, which is triggered by the matching, repositioning or serving passengers.

\noindent \textbf{Order generation}: The arrival of passengers and their destinations strictly follow the historical trip data as an input. Each passenger is associated with a maximum waiting time beyond which they will reject further services. 

\noindent \textbf{Matching module}: We conduct a long-waiting-prioritized greedy matching strategy, which sequentially assigns idle vehicles to the open orders with shortest customer waiting time. The idle vehicles include the vehicles in idling and cruising statuses. For the cruising vehicle, it is considered available at its real-time location. After being matched, it will stop cruising and move to the assigned pick-up location.

\noindent \textbf{Dispatching module}: The unmatched vehicles will next relocate to one of the six nearest hexagons. Specifically, if the agent selects the \textit{option}, it will stick to the primitive action that is generated by the option network until reaching the termination condition.

\section{Proof for Proposition~\ref{prop:position}}
\begin{proof}
Let $s_x$ and $s_y$ be two states of interest, and consider the Laplacian matrix $L$ of dimension $N \times N$ (with a total of $N$ states). Let $f(s_x),f(s_y)$ as the Laplacian representation obtained by minimizing Eqn.~\ref{eq:eig_draw_obj}. We consider the scenario where $||f(s_x)||<||f(s_y)||$ but there are more relocation trips associated with $s_y$ than $s_x$. The latter asserts that $L_{yy}>L_{xx}$. Given $u_1,...,u_D$ and $L$ being symmetric, individual term of the summation in Eqn.~\ref{eq:eig_draw_obj} can be expanded explicitly as:
\begin{equation}
    \textbf{u}_i^TL\textbf{u}_i= \sum_a L_{aa}u_{ia}^2+2\sum_{a,b,a\neq b} L_{ab}u_{ia}u_{ib}
\end{equation}
and
\begin{equation}
\begin{aligned}
    \sum_i \textbf{u}_i^TL\textbf{u}_i&=\sum_i(\sum_a L_{aa}u_{ia}^2+2\sum_{a,b,a\neq b} L_{ab}u_{ia}u_{ib})\\
    &=\sum_i(L_{xx}u_{ix}^2+L_{yy}u_{iy}^2)+C\\
    &=L_{xx}\sum_i u_{ix}^2+L_{yy}\sum_i u_{iy}^2 + C\\
    &=L_{xx}||f(s_x)||^2+L_{yy}||f(s_y)||^2+C
\end{aligned}
\end{equation}
Note that if we switch $f(s_x)$ with $f(s_y)$, the constant part $C$ will remain unchanged. On the other hand, we have:
\begin{equation}
    L_{xx}||f(s_y)||^2+L_{yy}||f(s_x)||^2 < L_{xx}||f(s_x)||^2+L_{yy}||f(s_y)||^2
\end{equation}

This poses a contradiction to the above scenario and this asserts that one can find a Laplacian representation that minimizes $\sum_i \textbf{u}_i^TL\textbf{u}_i$ such that $L_{yy}>L_{xx}$ and $||f(s_y)||<||f(s_x)||$ always hold. 
\end{proof}

\section{Parameter and network setting}
The training is performed under 5 different random seeds for each model using 14 days of weekday trip data. We set the learning rate to 0.001 for both high-level and low-level policy. The batch sizes are 256 for DQN, 128 for deep relocation policies and 32 for learning Laplacian embedding. We adopt the replay buffers of 100,000 samples for DQN, and 30,000 trajectories for deep relocation policies. the target network is updated every 100 ticks. The discount rate is set to $0.99$ for the high-level policy and $0.9$ for deep relocation policies. 

\begin{figure}[ht]
    \centering
    \subfloat[Neural network for training options]{\includegraphics[width=0.3\linewidth]{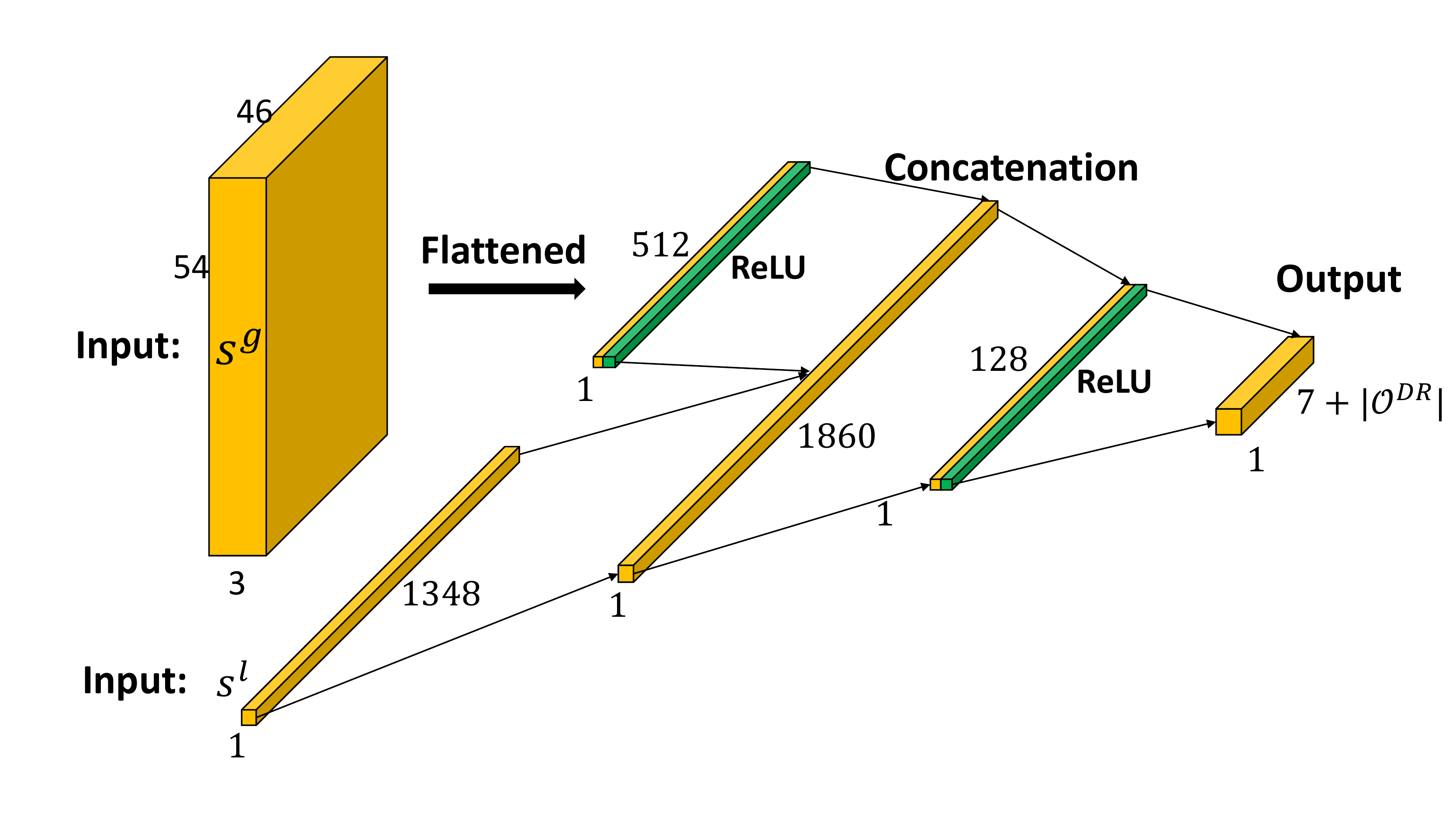}\label{fig:option_nn}}
    \subfloat[Neural network for training $f(\cdot)$ eigenvalues]{\includegraphics[width=0.3\linewidth]{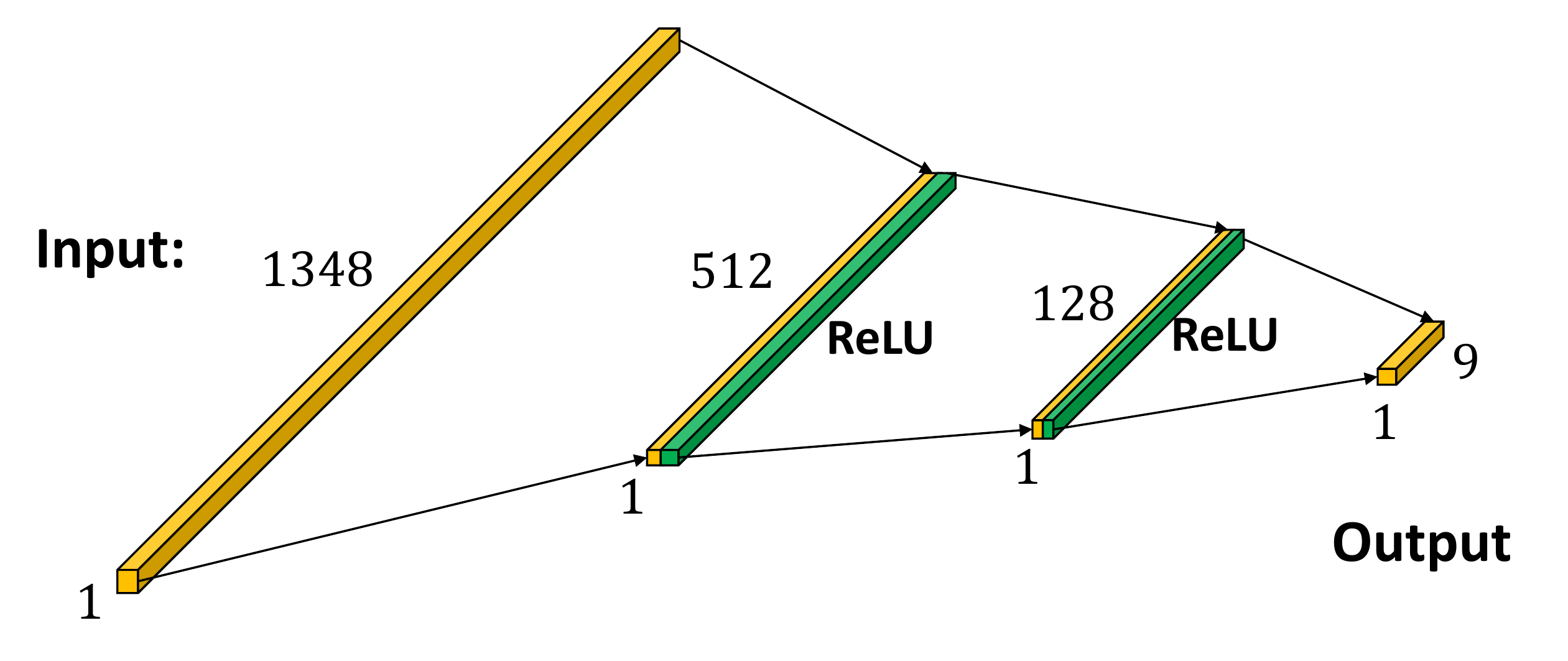}\label{fig:f_nn}}
    \caption{Illustration of neural networks used in this study}
    \label{fig:nns}
\end{figure}

In Figs.~\ref{fig:option_nn} and~\ref{fig:f_nn}, we present the architecture of the neural networks used in this study. To train the options, we use both global and local state representations ($\mathcal{S}^{g}$ and $\mathcal{S}^{l}$). The former is a $54\times46\times3$ tensor, then is flattened with ReLU activations. Following the concatenation with the input of $\mathcal{S}^{l}$, there is one fully connected layer with 128 hidden units with ReLU activations. Finally, the dimension of output depends on the number of primitive actions and the attached options, denoted by $7+|\mathcal{O}^{DR}|$. As in Fig.~\ref{fig:f_nn}, the network structure is consists of two fully connected layers with 512 and 128 hidden units with ReLU activations, and the resulted outputs are of $9$ dimensions. We use the last 8 dimensions of the output as the learned Laplacian representation for each state $s$.

% \section{Environment setting}
\end{document}